\documentstyle[12pt,aaspp4]{article}
\begin{document}
\baselineskip=18pt
\title{COMMENT ON ``PHOTON SPLITTING IN STRONGLY MAGNETIZED
 OBJECTS REVISITED''}

\author{Stephen L. Adler}
\affil{Institute for Advanced Study, 
 Princeton, NJ 08540}


\begin{abstract}
I point out that the results stated in the recent articles on photon 
splitting by Wunner, Sang, and Berg and by Wentzel, Berg, \& Wunner  
directly contradict an earlier analytic and numerical calculation that 
I performed of the same process using Schwinger's proper time method, 
for strong magnetic fields and general energies below the pair production 
threshold.  The results of Wunner et al. and Wentzel et al.  
do not show the expected low frequency behavior, nor have they been able 
to reproduce the expected small magnetic field behavior, suggesting that 
their calculations may not be gauge invariant.  
\end{abstract}

In a recent letter in this journal, Wunner, Sang, \& Berg (1995) 
have argued, on 
the basis of a detailed calculation of the photon splitting rate  
or absorption coefficient in an external magnetic field by 
Wentzel, Berg, \& Wunner (1994), that the photon splitting process has a 
much larger rate than was previously believed.  
Their article suggests that the large splitting rate results from  
inclusion 
of effects associated with magnetic fields $B$ of order $B_{cr} =4.4 \times 
10^{13} {\rm G}$, and photon energies of order the electron mass $m$, which 
they state had not been done in earlier calculations.  Wunner et al. 
correctly emphasize that their calculations, if correct, have important 
implications for cosmic $\gamma$ and x--ray sources.  I am writing this note 
to point out serious problems with the results of Wunner et al.  
which suggest that their numerical calculations may be in error (or may not 
be gauge invariant).  I begin by noting 
that Wunner et al. have made significant misstatements of fact  
in their references to the earlier literature, when they state that ``the 
astrophysical implications of magnetic photon splitting had to rely on 
simple analytical expressions derived by Adler (1971) and Papanyan \& Ritus
(1972) valid only in the weak-field limit $B << B_{cr}$...''.  This statement 
in fact applies only to the earlier letter by Adler, Bahcall, Callan,
\&  
Rosenbluth (1970) and {\it not} to the follow--up article of Adler (1971).   
In the Adler  et al. letter, the authors 
showed that gauge invariance implies that the leading
contribution to photon splitting comes from the hexagon diagram; they then  
calculated 
the contribution from this diagram to the photon splitting rate, and   
discussed its physical implications.  In the subsequent
Annals of Physics article of Adler (1971), I applied Schwinger's 
manifestly gauge invariant proper time method to give a compact expression 
for the photon splitting matrix element, valid for {\it arbitrarily large} 
magnetic field and for any photon energy below the pair production 
threshold.  (I used in this article the notation $\bar B$ for what I here 
term $B$.)  The matrix element expression (for the allowed polarization case) 
is given on pages 610--611
of the Annals article, and a graph showing the results of a numerical evaluation 
is given on page 613; a sketch of how the proper time calculation is 
performed is given in Appendix I on pages 634--644  (the full algebraic details 
of the photon splitting matrix element calculation amount to over 100 pages,
which I still retain in my files).  An important consistency check on the 
proper time calculation is that it reduces, in the weak field limit, to 
the hexagon diagram result calculated in the letter of Adler et al.   
This was checked both analytically and numerically; in fact the graph 
of the numerical work plots the ratio of the exact to leading order photon 
splitting rates or absorption coefficients, which 
approaches unity in the small 
magnetic field limit.  The numerical results show that for both $\omega=0$ 
and $\omega=m$, the ratio of the exact absorption coefficient to the hexagon 
expression is monotonically decreasing as $B$ increases from $0$ to $B_{cr}$,
and is only a weak function of $\omega$, in direct 
contradiction to the results 
obtained by Wunner, Sang, and Berg.

On examining the article of Wunner, Sang, and Berg and the calculation of 
Mentzel, Berg, and Wunner on which it is based, I am struck by the fact that
they never show, either analytically or numerically, that their photon 
splitting rate has the correct $B^6$ dependence for small $B$, 
nor do their 
numerical results show any evidence of the $\omega^5$ dependence 
expected for small values 
of $\omega/m$.  Wunner et al. attribute their inability to reproduce the  
leading order results to an anomalously low transition from the leading order 
behavior, stating ``Evidently at these field strengths 
the range of applicability 
of the weak-field, low-frequency form of the exact expression for photon 
splitting is restricted to much smaller photon energies than was previously 
thought''.  
However, there is no precedent for such anomalous 
behavior in any of the extensive  
calculations which have been performed in quantum electrodynamics.  
I have always considered it axiomatic, in performing a 
complicated analytic and numerical calculation, 
that results must be assumed to
be {\it wrong} unless one can reproduce one or more easily calculable 
limiting cases, and I find it  disturbing that this criterion has not been 
applied by Wunner et al.  I strongly 
suspect that the results obtained by 
these authors are incorrect because they have not maintained gauge invariance.  
I note that they have calculated in a particular gauge (Landau gauge), 
rather than working in a general gauge and using gauge invariance as a check 
on the manipulations.  This opens up the danger that any error or 
approximations which violate gauge invariance will introduce spurious 
contributions from terms of order $B$ in the amplitude, whereas 
these terms cancel by gauge invariance and the masslessness of the photon, 
with the leading contribution to the photon splitting amplitude coming 
in order $B^3$, with a coefficient proportional to the product $\omega  
\omega_1 \omega_2$ of the incoming and outgoing photon frequencies.

Because of the potential astrophysical implications of the high photon 
splitting absorption rate claimed by Wunner, Sang, and Berg, 
it is important that 
their calculation and mine be rechecked by a third party, with the aim 
of understanding where the discrepancy arises and determining who is right.   
I will be happy to send 
a copy of the full details of my analytic calculation, and my computer 
program notes and listing, to anyone wishing to perform this recalculation, 
and I trust that Wunner et al. will be willing to do the same.
\bigskip

This note is based on a letter which I wrote to  Drs. Wunner, Sang, 
and Berg in April, 1995, to which I received no response.  
I wish to thank John Bahcall and Bohdan 
Paczynski for urging that the issues be aired in a public forum.  This 
work was supported in part by the Department of Energy under Grant 
\#DE-FG02-90ER40542.


\begin{thebibliography}{}
\bibitem{}Adler, S. L., Bahcall, J. N., Callan, C. G., \& Rosenbluth, M. N. 1970, 
Phys. Rev. Lett. 25, 1061
\bibitem{}Adler, S. L. 1971, Ann. Phys. 67, 599
\bibitem{}Mentzel, M., Berg, D., \& Wunner, G. 1994.  Phys. Rev. D, 50, 1125
\bibitem{}Wunner, G., Sang, R., \& Berg, D. 1995, ApJ 455, L51
\end{thebibliography}
\end{document}